\documentstyle[aps]{revtex}
\begin{document}
\title{{\bf Probabilistic Teleportation and Entanglement Matching}}
\author{Wan-Li Li, Chuan-Feng Li, Guang-Can Guo\thanks{%
Email: gcguo@ustc.edu.cn}}
\address{Laboratory of Quantum Communication and Quantum Computation and Department\\
of Physics, University of Science and Technology of China,\\
Hefei, 230026, P. R. China\vspace{0.5in}}
\maketitle

\begin{abstract}
\baselineskip12ptTeleportation may be taken as sending and extracting
quantum information through quantum channels. In this report, it is shown
that to get the maximal probability of exact teleportation through partially
entangled quantum channels, the sender (Alice) need only to operate a
measurement which satisfy an ``entanglement matching'' to this channel. An
optimal strategy is also provided for the receiver (Bob) to extract the
quantum information by adopting general evolutions.

PACS Number(s) 03.67$^{*}$, 03.65.Bz
\end{abstract}

\baselineskip12ptQuantum teleportation, the process that transmits an
unknown qubit from a sender (Alice) to a receiver (Bob) via a quantum
channel with the help of some classical information, is originally concerned
by Bennett, Brassard, Crepeau, Jozsa, Peres, and Wootters (BBCJPW) $\left[
1\right] $. In their scheme, such a quantum channel is represented by a
maximally entangled pair (any of the Bell states) and the original state
could be deterministically transmitted to Bob.

The process of teleportation may be regarded as sending and extracting
quantum information via the quantum channel. We will apply this picture to
investigate a partially entangled quantum channel. Because a mixed state can
be purified to a Bell state with no probability $\left[ 2-4\right] $, a
quantum channel of mixed state could never provide an teleportation with
fidelity 1. Therefore only pure entangled pairs should be considered if we
prefer an exact teleportation (even with some probability). For the reason
of Schmidt disposition $\left[ 5\right] $, a partially entangled pair may be
expressed as 
\begin{equation}
\left| \Phi \right\rangle _{2,3}=a\left| 00\right\rangle _{2,3}+b\left|
11\right\rangle _{2,3}\text{ }(\left| a\right| ^2+\left| b\right| ^2=1,\text{
}\left| a\right| >\left| b\right| ).
\end{equation}
(hereafter, we assume particle 2 is at Alice's site and particle 3 at Bob's
site) Absolute value of the Schmidt coefficient $\left| b\right| $ is an
invariant under local operations, and it corresponds to the entanglement
entropy $E$ of the state $\left[ 6\right] $. Such a state can be
concentrated to a Bell state $\left[ 6,7\right] $ with the probability of $%
2b^2$ and the concentrated pair may be used as a new quantum channel to
carry out a teleportation.

In this report, Alice performs a Von-Neumann measurement on her side while
Bob performs a corresponding general evolution to reestablish the initial
state with a certain probability. We will give a measure of the entanglement
degree to Alice's measurement and show that the optimal probability of an
exact teleportation is determined by the less one of the entanglement
degrees of Alice's measurement and the quantum channel. Thus the matching of
these entanglement degrees should be considered and the entanglement degree
of the measurement is endowed a meaning of Alice's ability to send quantum
information.

First, we consider the case Alice operates a Bell measurement and give Bob's
proper general evolution to reestablish the initial state with an optimal
probability. Considering the previously shared pair shown in Eq. $\left(
1\right) $ and the unknown state (which is to be send) of particle 1 $\mid
\phi \rangle _1=\alpha \left| 0\right\rangle _1+\beta \left| 1\right\rangle
_1$, the total state could be written as $\left| \Psi \right\rangle
_{1,2,3}=\left| \phi \right\rangle _1\left| \Phi \right\rangle _{2,3}=\alpha
a\left| 000\right\rangle _{1,2,3}+\alpha b\left| 011\right\rangle
_{1,2,3}+\beta a\left| 100\right\rangle _{1,2,3}+\beta b\left|
111\right\rangle _{1,2,3}$. If Alice operates a Bell measurement, Bob will
get the corresponding unnomalized states as the following: 
\begin{equation}
\left. 
\begin{array}{c}
\langle \Phi _{1,2}^{+}\mid \Psi \rangle _{1,2,3}=\frac{\sqrt{2}}2\left(
\alpha a\left| 0\right\rangle _3+\beta b\left| 1\right\rangle _3\right) , \\ 
\langle \Phi _{1,2}^{-}\mid \Psi \rangle _{1,2,3}=\frac{\sqrt{2}}2\left(
\alpha a\left| 0\right\rangle _3-\beta b\left| 1\right\rangle _3\right) , \\ 
\langle \Psi _{1,2}^{+}\mid \Psi \rangle _{1,2,3}=\frac{\sqrt{2}}2\left(
\beta a\left| 0\right\rangle _3+\alpha b\left| 1\right\rangle _3\right) , \\ 
\langle \Psi _{1,2}^{-}\mid \Psi \rangle _{1,2,3}=\frac{\sqrt{2}}2\left(
\beta a\left| 0\right\rangle _3-\alpha b\left| 1\right\rangle _3\right) .
\end{array}
\right.
\end{equation}
where $\left\{ \left| \Phi _{1,2}^{\pm }\right\rangle =\frac{\sqrt{2}}2%
\left( \left| 00\right\rangle _{1,2}\pm \left| 11\right\rangle _{1,2}\right)
,\text{ }\left| \Psi _{1,2}^{\pm }\right\rangle =\frac{\sqrt{2}}2\left(
\left| 01\right\rangle _{1,2}\pm \left| 10\right\rangle _{1,2}\right)
\right\} $ are Bell states of particle 1 and particle 2. Alice informs Bob
her measurement result, for example $\left| \Phi _{1,2}^{+}\right\rangle $
(with the corresponding collapsed state of particle 3 as $\langle \Phi
_{1,2}^{+}\mid \Psi \rangle _{1,2,3}=\frac{\sqrt{2}}2\left( \alpha a\left|
0\right\rangle _3+\beta b\left| 1\right\rangle _3\right) $ which is
unnomalized), and Bob gives a corresponding general evolution. To carry out
a general evolution, an auxiliary qubit with the original state $\left|
0\right\rangle _{aux}$ is introduced. Under the basis $\left\{ \left|
0\right\rangle _3\left| 0\right\rangle _{aux},\left| 1\right\rangle _3\left|
0\right\rangle _{aux},\left| 0\right\rangle _3\left| 1\right\rangle
_{aux},\left| 1\right\rangle _3\left| 1\right\rangle _{aux}\right\} $, a
collective unitary transformation 
\begin{equation}
U_{sim}=\left( 
\begin{array}{cccc}
\frac ba & 0 & \sqrt{1-\frac{b^2}{a^2}} & 0 \\ 
0 & 1 & 0 & 0 \\ 
0 & 0 & 0 & -1 \\ 
\sqrt{1-\frac{b^2}{a^2}} & 0 & -\frac ba & 0
\end{array}
\right) ,
\end{equation}
transforms the unnomalized product state $\frac{\sqrt{2}}2\left( \alpha
a\left| 0\right\rangle _3\left| 0\right\rangle _{aux}+\beta b\left|
1\right\rangle _3\left| 0\right\rangle _{aux}\right) $ to the result: 
\begin{equation}
\left| \Phi \right\rangle _{3,aux}=\frac{\sqrt{2}}2\left[ b\left( \alpha
\left| 0\right\rangle _3+\beta \left| 1\right\rangle _3\right) \left|
0\right\rangle _{aux}+a\sqrt{1-\frac{b^2}{a^2}}\alpha \left| 1\right\rangle
_3\left| 1\right\rangle _{aux}\right] ,
\end{equation}
which is also unnormalized. Then a measurement to the auxiliary particle
follows. If the measurement result is $\left| 0\right\rangle _{aux}$, the
teleportation is successfully accessed, while if the result is $\left|
1\right\rangle _{aux}$, teleportation fails with the state of qubit 3
transformed to a blank state $\left| 1\right\rangle _3$ and no information
about the initial qubit 1 left (thus an optimal probability of teleportation
is accessed). The contribution of this unnomalized state to the probability
of successful teleportation may be expressed by the probabilistic amplitude
of $\alpha \left| 0\right\rangle _3+\beta \left| 1\right\rangle _3$ in Eq. $%
\left( 4\right) $ as $\left| (\frac{\sqrt{2}}2b)\right| ^2=\frac 12\left|
b\right| ^2$.

Other states in Eq. $\left( 2\right) $ could be discussed in the same way,
and their contributions to the probability of successful teleportation may
be calculated directly by using a general method: if the unnormalized state
in Eq. $\left( 2\right) $ is written as $\alpha x\left| 0\right\rangle
_3+\beta y\left| 1\right\rangle _3$ or $\alpha x\left| 1\right\rangle
_3+\beta y\left| 0\right\rangle _3$, after Bob's optimal operation, it gives
a contribution to the whole successful probability as 
\begin{equation}
p=\left( \min \left\{ \left| x\right| ,\left| y\right| \right\} \right) ^2.
\end{equation}
Adding up all the contributions, the optimal probability of successful
teleportation is obtained as $P=\frac 12\left| b\right| ^2\times 4=2\left|
b\right| ^2$.

Then consider more general cases: Alice operates a measurement with such
eigenstates: 
\begin{equation}
\left. 
\begin{array}{c}
\left| \Phi \right\rangle _{1,2}^1=a^{^{\prime }}\left| 00\right\rangle
_{1,2}+b^{^{\prime }}\left| 11\right\rangle _{1,2}, \\ 
\left| \Phi \right\rangle _{1,2}^2=b^{^{\prime }}\left| 00\right\rangle
_{1,2}-a^{^{\prime }}\left| 11\right\rangle _{1,2}, \\ 
\left| \Phi \right\rangle _{1,2}^3=a^{^{\prime }}\left| 10\right\rangle
_{1,2}+b^{^{\prime }}\left| 01\right\rangle _{1,2}, \\ 
\left| \Phi \right\rangle _{1,2}^4=b^{^{\prime }}\left| 10\right\rangle
_{1,2}-a^{\prime }\left| 01\right\rangle _{1,2}.
\end{array}
\right. (\left| a^{^{\prime }}\right| ^2+\left| b^{^{\prime }}\right|
^2=1,\left| a^{^{\prime }}\right| \geq \left| b^{^{\prime }}\right| ).
\end{equation}
For the reason of Schimidt disposition, this basis has represented all
possible Von-Neumann measurements of two particles when $\left( a^{^{\prime
}},b^{^{\prime }}\right) $ varies. The four states above are orthogonal and
have the same entanglement entropy, so the measurement's entanglement degree 
$E$ can be defined as that of any of the four states. Collapsed states of
particle 3 corresponding to the four measurement result could be written as: 
\begin{equation}
\left. 
\begin{array}{c}
\langle \Phi _{1,2}^1\mid \Psi \rangle _{1,2,3}=\alpha aa^{^{\prime }}\left|
0\right\rangle _3+\beta bb^{^{\prime }}\left| 1\right\rangle _3, \\ 
\langle \Phi _{1,2}^2\mid \Psi \rangle _{1,2,3}=\alpha ab^{^{\prime }}\left|
0\right\rangle _3-\beta ba^{^{\prime }}\left| 1\right\rangle _3, \\ 
\langle \Phi _{1,2}^3\mid \Psi \rangle _{1,2,3}=\beta aa^{^{\prime }}\left|
0\right\rangle _3+\alpha bb^{^{\prime }}\left| 1\right\rangle _3, \\ 
\langle \Phi _{1,2}^4\mid \Psi \rangle _{1,2,3}=\beta ab^{^{\prime }}\left|
0\right\rangle _3-\alpha ba^{^{\prime }}\left| 1\right\rangle _3,
\end{array}
\right.
\end{equation}
which is unnormalized. The general evolution to particle 3 is similar to
what is shown in Eq. $\left( 3\right) $. From the result of Eq. $\left(
6\right) $, the probability of successful teleportation could be considered
directly in the following two cases:

1. $\left| a\right| \geq \left| a^{^{\prime }}\right| \geq \left|
b^{^{\prime }}\right| \geq \left| b\right| $

In this case, because $\left| \left( ab^{^{\prime }}\right) \right|
^2=\left| a\right| ^2$ $\left( 1-\left| a^{^{\prime }}\right| ^2\right) $
and $\left| \left( ba^{^{\prime }}\right) \right| ^2=\left| a^{^{\prime
}}\right| ^2\left( 1-\left| a\right| ^2\right) $, inequality $\left|
ab^{^{\prime }}\right| \geq \left| ba^{^{\prime }}\right| $ is established,
and $\left| aa^{^{\prime }}\right| \geq \left| bb^{^{\prime }}\right| $ is
obvious, so the whole probability of successful teleportation may be written
as 
\[
P=\left| \left( bb^{^{\prime }}\right) \right| ^2+\left| \left( ba^{^{\prime
}}\right) \right| ^2+\left| \left( bb^{^{\prime }}\right) \right| ^2+\left|
\left( ba^{^{\prime }}\right) \right| ^2=2\left| b\right| ^2, 
\]
which is just the same as the case Alice operates a Bell measurement.

2. $\left| a^{^{\prime }}\right| \geq \left| a\right| \geq \left| b\right|
\geq \left| b^{^{\prime }}\right| $

In this case, $\left| ba^{^{\prime }}\right| \geq \left| ab^{^{\prime
}}\right| $, and the probability of successful teleportation is 
\begin{equation}
P=\left| \left( bb^{^{\prime }}\right) \right| ^2+\left| \left( ab^{^{\prime
}}\right) \right| ^2+\left| \left( bb^{^{\prime }}\right) \right| ^2+\left|
\left( ab^{^{\prime }}\right) \right| ^2=2\left| b^{^{\prime }}\right| ^2.
\end{equation}

From the analysis above, the probability of successful teleportation is
determined by the less one of $\left| b\right| $ and $\left| b^{^{\prime
}}\right| $, and may be regarded as being determined by the less
entanglement degree of Alice's measurement and the quantum channel.

Just as what is mentioned above, teleportation may be regarded as the
quantum channel's preparation and quantum information's sending and
extraction. The result above may be explained clearly by using this picture.
The entanglement degree of Alice's measurement could be considered as
Alice's sending ability and the entanglement degree of the quantum channel
could be taken as the width of it. Then the amount of transmitted quantum
information is determined by the lower one of these two bounds: the width of
the quantum channel $2\left| b\right| ^2$ and the sending ability of Alice $%
2\left| b^{^{\prime }}\right| ^2$. If they are just the same, an
``entanglement matching'' is satisfied. If Bob always reestablish the
to-be-sent state in an optimal probability (which means he always extract
all the quantum information he received), an exact teleportation will be
performed with the probability equal to the amount of the quantum
information transmitted, just as what is shown in Eqs. $\left( 8,9\right) $.

Though Bell measurement is an essential task of quantum teleportation, it is
very difficult to be fully accessed and It has been shown that Bell states
cannot be distinguished completely by using linear devices $\left[
8,9\right] $, while this difficulty can be seen in some teleportation
experiments $\left[ 10\right] $. Von-Neumann Measurements with less
entangled eigenstates may be more efficient. From the result above, if a
partial entanglement state $\left| \Phi \right\rangle _{2,3}=a\left|
00\right\rangle _{2,3}+b\left| 11\right\rangle _{2,3}$ is adopted as the
quantum channel, the same optimal probability of successful teleportation
could be accessed if only Alice's measurement satisfied the ``entanglement
matching'', while a Bell measurement or a POVM is not necessary. The
matching here is essential to get an optimal probability, and it could also
be regarded as the matching between the quantum channel's width and Alice's
sending ability. Without such a matching, a waste of quantum information
either at Alice's site or through the quantum channel will be caused.

The result of entanglement matching can be generalized to the teleportation
of multi-particle system. Considering a $k$-particle system $P$ at Alice's
site with the state $\left| \Psi \right\rangle _P=$ $\alpha _0\left| 00\cdot
\cdot \cdot 00\right\rangle _{P_1,\cdot \cdot \cdot ,P_k}+\alpha _1\left|
00\cdot \cdot \cdot 01\right\rangle _{P_1,\cdot \cdot \cdot ,P_k}+\cdot
\cdot \cdot \cdot \cdot \cdot +\alpha _{2^k-1}\left| 11\cdot \cdot \cdot
11\right\rangle _{P_1,\cdot \cdot \cdot ,P_k}$. Without loss of generality,
the quantum channel between Alice and Bob is $k$ independent entangled pairs
with the state $\prod\limits_{i=1}^k\left( a_i\left| 00\right\rangle
_{A_i,B_i}+b_i\left| 11\right\rangle _{A_i,B_i}\right) $ (any other pure
quantum channel could be transformed to this by local operations). Alice
draws $k$ collective measurements, each of which is Von-Neumann measurement
with the following eigenstates: 
\begin{equation}
\left. 
\begin{array}{c}
\left| \Phi \right\rangle ^{i,1}=a_i^{^{\prime }}\left| 00\right\rangle
_{P_i,A_i}+b_i^{^{\prime }}\left| 11\right\rangle _{P_i,A_i}, \\ 
\left| \Phi \right\rangle ^{i,2}=b_i^{^{\prime }}\left| 00\right\rangle
_{P_i,A_i}-a_i^{^{\prime }}\left| 11\right\rangle _{P_i,A_i}, \\ 
\left| \Phi \right\rangle ^{i,3}=a_i^{^{\prime }}\left| 10\right\rangle
_{P_i,A_i}+b_i^{^{\prime }}\left| 01\right\rangle _{P_i,A_i}, \\ 
\left| \Phi \right\rangle ^{i,4}=b_i^{^{\prime }}\left| 10\right\rangle
_{P_i,A_i}-a_i^{\prime }\left| 01\right\rangle _{P_i,A_i}.
\end{array}
\right. (\left| a_i^{^{\prime }}\right| ^2+\left| b_i^{^{\prime }}\right|
^2=1,\text{ }\left| a_i^{^{\prime }}\right| \geq \left| b_i^{^{\prime
}}\right| ),
\end{equation}
where $i=1,2,\cdot \cdot \cdot ,k$. Then Bob reestablishes the original
state as $\left| \Psi \right\rangle ^B$ with a certain probability by
adopting a proper general evolution. Using similar methods as the case of
mono-qubit teleportation, we may show that there also exists an entanglement
matching in multi-qubit teleportation: If $c_i$ is defined as $\min \left\{
\left| b_i^{^{\prime }}\right| ,\left| b_i\right| \right\} $, the optimal
probability of successful teleportation could be expressed as $%
2^k\prod\limits_{i=1}^kc_i^2$.

This work was supported by the National Natural Science Foundation of China.

\end{document}